\documentclass[twocolumn,english,aps,pra]{revtex4-1}
\usepackage{comment,amsmath,graphicx,amssymb,epsfig,babel,dsfont,color,subfigure}
\usepackage[latin1]{inputenc}
\usepackage{textcomp}

\renewcommand{\Im}{{\rm Im}}

\newcommand{\rd}{{\rm d}}

\newcommand{\ri}{{\rm i}}

\begin{document}

\title{Strong Coupling of Collection of Emitters on Hyperbolic Meta-Material}

\author{Svend-Age Biehs}
\affiliation{Institut f\"{u}r Physik, Carl von Ossietzky Universit\"{a}t, D-26111 Oldenburg, Germany}

\author{Chenran Xu}
\affiliation{MOE Key Laboratory of Advanced Micro-Structured Materials, School of Physics Science and
Engineering, Tongji University, Shanghai 200092, China}

\author{Girish S. Agarwal}
\affiliation{ Institute for Quantum Science and Engineering and Department of Biological and 
Agricultural Engineering Department of Physics and Astronomy, Texas A \& M University, College Station, Texas 77845, USA}

\email{s.age.biehs@uni-oldenburg.de} 

\begin{abstract}
Recently considerable effort is devoted to the realization of the strong coupling regime of the radiation matter interaction in 
the context of emitter at a meta surface. The strong interaction is well realized in cavity quantum electrodynamics which also 
shows that the strong coupling is much easier to realize using a collection of emitters. Keeping this in mind we study if emitters 
on a hyperbolic meta  materials can yield strong coupling regime. We show that the strong coupling can be realized for densities 
of emitters exceeding a critical value. A way to detect the strong coupling between emitters and a hyperbolic metamaterials is to 
use the Kretschman-Raether configuration. The strong coupling appears as the splitting of the reflectivity dip. In  the weak coupling 
regime the dip position shifts. The shift and splitting can be used to sense active molecules at surfaces.  
\end{abstract}

\maketitle

\section{Introduction}

It is well known that the emission rate of atoms and molecules can be altered by changing the environment 
of the emitters~\cite{Drexhage1974,Agarwal1975,Barnes1}. When bringing such emitters close to a plasmonic 
or hyperbolic film, the emission rate can be changed drastically due to the coupling with the surface plasmons 
polaritons or hyperbolic modes~\cite{Agarwal1975,Barnes1,AgarwalVollmer,Barnes,ZubinEtAl2012,KidwaiEtAl2012,KimEtAl2012,KrishnamorthyEtAl2012,GalfskyEtAl2015,WangEtAl2015}. While the emission rate is 
increased considerably the question remains --- can one reach the strong coupling regime well known from 
cavity QED. For a single emitter on  a plasmonic surface this is almost impossible to reach, as we argue 
in Sec II. This is due to the very large damping of the localized or surface plasmons. However it is known~\cite{Agarwal1984} 
that a collection of emitters effectively enhances the coupling and thus collective excitations of the emitters can be 
used to reach the strong coupling regime. The usefulness of collective excitations has been demonstrated for a very wide variety 
of systems: superconducting qubits~\cite{FinkEtAl2009}; magnons~\cite{TabuchiEtAl2014,AbdurakhimovEtAl2015}; 
cyclotron motion of surface electrons~\cite{AbdurakhimovEtAl2016}; graphene ribbons~\cite{KimEtAl2016}; 
porphyrin excitons~\cite{BerrierEtAl2011}, J-aggregates~\cite{BelessaEtAl2004,SugawaraEtAl2006,WurtzEtAl2007,FofangEtAl2008,VasaEtAl2010,VasaEtAl2013}, dye molecules~\cite{Hakala2009,Vaekevainen2014}, quantum dots~\cite{GomezEtAl2010,GomezEtAl2010b} and nanoprisms~\cite{BalciEtAl2015}. Recently also strong coupling for monolayer transition dichalcogenides in a dielectric cavity~\cite{LiuEtAl2017} and in a plasmonic cavity~\cite{KleemannEtAl2017} has been realized.

In the last decade, hyperbolic materials~\cite{HuChui2002,Smith2003} have been shown to be a versatile platform allowing for 
broadband enhanced spontaneous emission~\cite{ZubinEtAl2012,KidwaiEtAl2012,KimEtAl2012,KrishnamorthyEtAl2012,GalfskyEtAl2015,WangEtAl2015}, hyperbolic lensing~\cite{JacobEtAl2006,FengAndElson2006,CegliaEtAl2014,Ben2012,Conteno2015}, negative refraction~\cite{SmithEtAl2004,HoffmanEtAl2007}, broadband enhanced thermal emission~\cite{Nefedov2011, Biehs2012,GuoEtAl2012,Biehs2,ShiEtAl2015,Biehs2015}, {near-field thermophotovoltaics~\cite{SimovskiEtAl2013,MirmoosaEtAl2017}}, long-range heat and energy transfer~\cite{MessinaEtAl2016,SABGSA2015,DesmukhEtAl2017}, {thermal and nonthermal nanoantennas~\cite{ValagiannopoulosEtAl2014,BarbillonEtAl2017}} as well as strong coupling of emitters embedded in such materials~\cite{Shekhar2014}. {Reviews dealing with wave propagation in hyperbolic metamaterials and applications are given in Refs.~\cite{PoddubnyEtAl2013,FerrariEtal2015,BoardmanEtAl2017}.} In this work we investigate if HMM could provide a good platform for realizing the strong coupling regime using the collective excitations of emitters. We show how the Kretchmann-Raether~\cite{Hakala2009,Vaekevainen2014} configuration can be used for investigations on strong coupling on HMM.

Our work is organized as follows: In section II we show what prevents the observation of strong coupling for a single emitter and 
what is required for seeing the effects of strong coupling. In section III we introduce the Kretchmann-Raether configuration,
the hyperbolic material used in that configuration and the material properties of the dye molecules. The Kretchmann-Raether 
configuration is the simplest way in which the features of the strong coupling can be explored. Finally, in section IV we discuss 
the transition from the weak to the strong-coupling regime by studying the reflectivity in the Kretchmann-Raether setup as a function 
of the concentration of the dye molecules. In the weak coupling regime the dip of the reflectivity shifts whereas in the strong 
coupling regime the reflectivity dip splits in to two. These features can be used as sensors of molecules on 
surfaces~\cite{ZhuEtAl2010,HeEtAl2011}. We summarize our main findings in section V.

\section{Rabi Splitting for a single Atom}

First, let us dwell on the possibility to detect a single atom by means of Rabi 
splitting. To this end, we consider a single atom at position $\mathbf{r}_A$ close to 
the surface of a substrate. Here, we describe the atom by a classical dipole because 
for Rabi splitting it is not essential to make a quantum approach. A nice comparison
of the classical and quantum approach can be found in Ref.~\cite{Barnes}. In the classical
case the dipole moment can be described as a harmonic oscillator obeying the 
standard equation of motion
\begin{equation}
  \biggl( \frac{\rd^2}{\rd t^2} + \omega_0^2 \biggr) \mathbf{p} = \frac{e^2}{m} \bigl( \mathbf{E}_{\rm ext} + \mathbf{E}_{\rm dip} \bigr)
\end{equation}
where $\mathbf{p} = e \mathbf{x}$, and $\mathbf{E}_{\rm ext}$ is the external field exciting the
dipole moment of the atom and $\mathbf{E}_{\rm dip}$ is the field generated by the dipole moment;
$\omega_0$ is the atomic transition frequency. In Fourier space it can be written as
\begin{equation}
  \mathbf{E}_{\rm dip} = \frac{\omega^2}{c^2 \epsilon_0} \bigl( \mathds{G}^{(vac)}(\mathbf{r}_A,\mathbf{r}_A;\omega) +  \mathds{G}^{(sc)}(\mathbf{r}_A,\mathbf{r}_A;\omega) \bigr)\cdot \mathbf{p}.
\end{equation}
introducing the vacuum and the scattering part $\mathds{G}^{(vac)}$ and $\mathds{G}^{(sc)}$ 
of the dyadic Green function evaluated at the position of the atom. By going into Fourier space and
by inserting the Green's function the equation of motion reads
\begin{equation}
  \mathds{A} \cdot \mathbf{p} = \frac{e^2}{m} \mathbf{E}_{\rm ext}
\label{Eq:EOM}
\end{equation}
with 
\begin{equation}
\begin{split}
    \mathds{A} &:= (\omega^2_0 - \omega^2) \mathds{1} - \ri \frac{\omega^2 e^2}{c^2 \epsilon_0 m} \Im \mathds{G}^{(vac)}(\mathbf{r}_A,\mathbf{r}_A;\omega) \\
               &\qquad - \frac{\omega^2 e^2}{c^2 \epsilon_0 m} \mathds{G}^{(sc)}(\mathbf{r}_A,\mathbf{r}_A;\omega) .
\end{split}
\end{equation}
Here we have neglected the divergent real part of the vacuum Green's function. This corresponds
to a renormalization of mass and frequency of the harmonic oscillator. 

In order to determine the eigen frequencies of the coupled field-atom system we need to solve Eq.~(\ref{Eq:EOM}). 
But it suffices to focus on the 'zeroes' of $\mathds{A}$ in order to find the eigen frequencies.
If we focus for example on the $zz$ component of $\mathds{A}$ only, then 
\begin{equation}
    A_{zz} := (\omega^2_0 - \omega^2) - 2 \ri  \gamma_{zz}^{(vac)} \omega - \frac{\omega^2 e^2}{c^2 \epsilon_0 m} \mathds{G}^{(sc)}_{zz}(\mathbf{r}_A,\mathbf{r}_A;\omega).
\end{equation}
Here 
\begin{equation}
  \gamma_{zz}^{(vac)} := \frac{\omega e^2}{2 c^2 \epsilon_0 m} \Im \mathds{G}^{(vac)}_{zz}(\mathbf{r}_A,\mathbf{r}_A;\omega)
\end{equation}
is essentially the emission rate of the atom in vacuum. The coupling constant of the atom-surface system can now be 
determined by assuming that the substrate itself has a surface mode resonance.  Around this resonance we can 
approximate the zz component of the scattered part of the Green's function as
\begin{equation}
  G_{zz}^{(sc)}(\omega) \approx \frac{- \chi}{\omega - \omega_c + \ri \kappa_c}.
\end{equation}
Here, $\chi$ is the oscillator strength and $\kappa_c$ is the damping constant of that surface resonance.
Assuming further that the resonance frequency  and that the excitation frequency of the atom
are close to resonance, i.e.\ $\omega_c \approx \omega_0$ and $\omega \approx \omega_0$ we can determine the
zeroes of $A_{zz}$ easily. For the eigen frequencies of the coupled system we then obtain
\begin{equation}
  \omega = \omega_0 - \ri \frac{\gamma_{zz}^{(vac)} + \kappa_c}{2} \pm \sqrt{g^2 - \frac{(\gamma_{zz}^{(vac)} - \kappa_c)^2}{4}}
\end{equation}
where we have introduced the coupling constant
\begin{equation}
  g^2 = \frac{e^2 \omega_0}{2 m c^2 \epsilon_0} \chi.
\end{equation}
Obviously, we have a damping or energy loss due to the spontaneaus emission into vacuum $\gamma_{zz}^{(vac)}$ 
and due to the losses in the surfaces wave. In the weak-coupling regime, i.e.\ for  $g^2 \ll (\gamma_{zz}^{(vac)} - \kappa_c)^2/4$  
the eigen frequencies of the coupled system coincide with $\omega_0$. On the other hand, for in the
strong coupling regime when $g \gg \kappa_c/2$ and $g \gg \gamma_{zz}^{(vac)}/2$ the degeneracy of the eigen 
frequencies of the coupled system is lifted and one can observe two distinct eigen frequencies which differ from $\omega_0$. 
Therefore, in order to observe this Rabi splitting it is necessary to enter the strong coupling regime.
If the atom is placed in the near-field of the surface mode resonance, then the condition $g \gg \gamma_{zz}^{(vac)}/2$ is 
automatically fullfilled due to the Purcell effect. On the other hand the condition
\begin{equation}
  \frac{2 g}{\kappa_c} \gg 1
\label{Eq:strongcoupling}
\end{equation}
is much harder to fulfill, because the damping of the surface waves becomes large close to the surface mode resonance. 
We have checked by numerical calculations that indeed this condition is hard to meet so that Rabi splitting for a single atom 
seems to be quite difficult, although for the coupling of single atoms to metal nanoparticles a strong-coupling is theoretically
possible~\cite{TrueglerEtAl2008}. However, it is known that the Rabi splitting scales like $\sqrt{N}$ where $N$ is the number of 
atoms~\cite{Agarwal1984}. Therefore by increasing the number of atoms, Rabi splitting becomes at some point measurable.

In view of this it is better to consider a dense medium whose linear response to electromagnetic fields can be described by the dielectric function
\begin{equation} 
  \epsilon(\omega) = 1 + \frac{\omega_p^2}{\omega_0^2 - \omega^2 - \ri \omega \gamma}.
\label{Eq:PermittivityMolecule}
\end{equation}
We further assume that such a medium is in a plasmonic cavity with losses described by $\kappa_c$. The cavity mode is tuned such that $\omega_c \approx \omega_0$. Then one finds in the strong coupling regime with $\omega_c \gg \gamma, \kappa_c$ the simple relation
\begin{equation}
  \omega \approx \omega_c \pm \frac{\omega_p}{2}
\label{Eq:Estimate}
\end{equation}
for the new coupled mode frequencies if $\omega_p \ll \omega_c$ and $\omega_p \gg \gamma, \kappa_c$. First, we note that the Rabi splitting is proportional
to $\omega_p$ and therefore proportional to $\sqrt{N}$ as expected~\cite{Agarwal1984,Barnes}. To see the Rabi splitting the density of atoms described 
by $\omega_p$ has to be larger than the losses of the plasmonic cavity $\kappa_c$. This is in accordance with our previous result in (\ref{Eq:strongcoupling}) and makes clear that due to the losses of the metall a large density of atoms is necessary to see Rabi splitting. {Note that this gives an estimate of the new normal modes. Actual values are to be obtained from solutions of full Maxwell equations for the anisotropic medium.}

\begin{figure}[!htb]
   \includegraphics[width=0.45\textwidth]{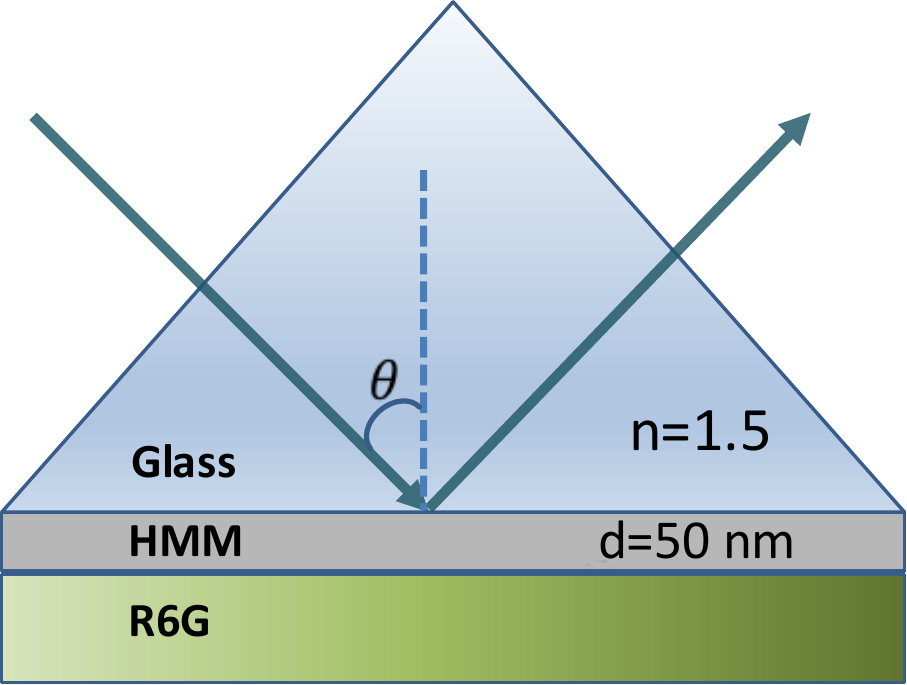}
   \caption{Sketch of the Kretschmann configuration for Rabi-splitting measurement. \label{Fig:Kretschmann}}
\end{figure}

\section{Kretschmann-Raether setup}

\begin{figure}[!htb]
   \includegraphics[width=0.45\textwidth]{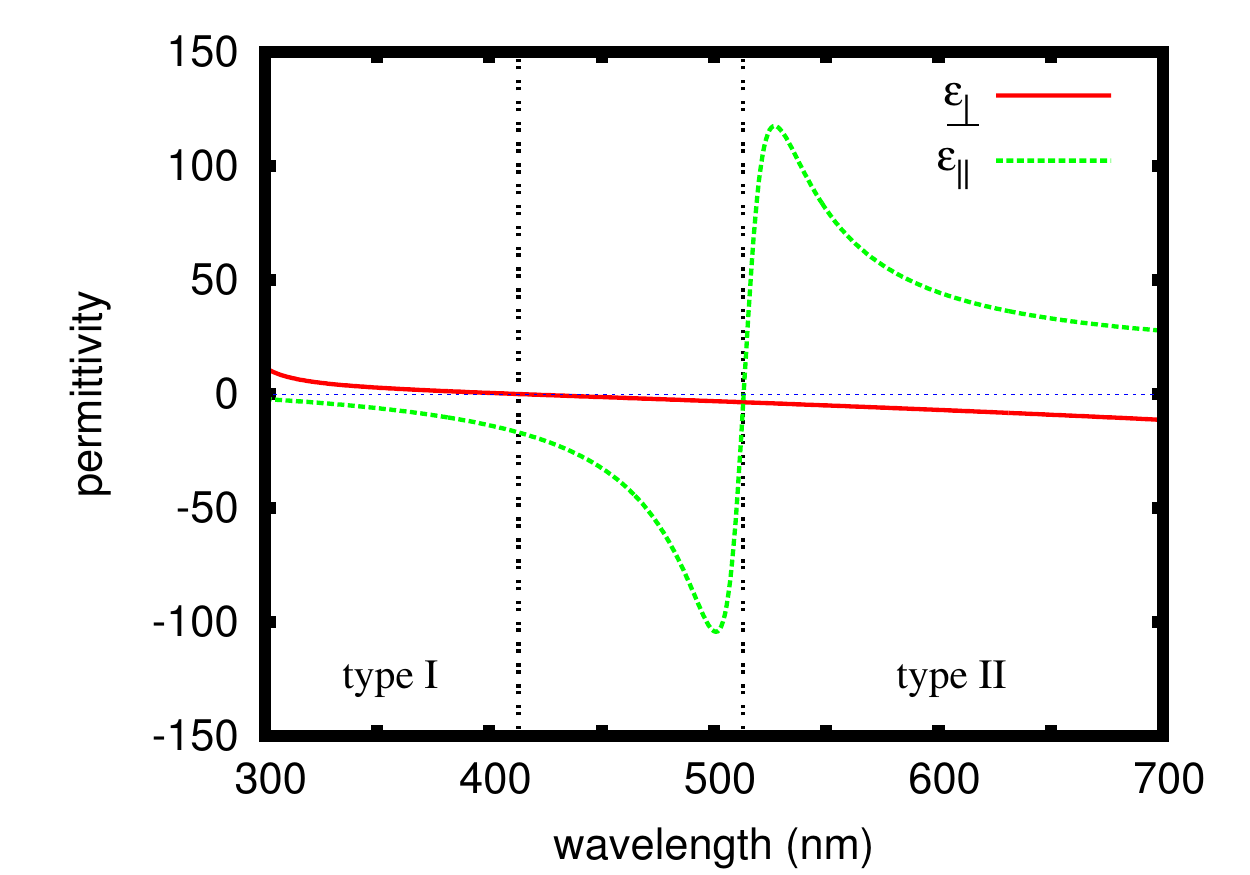}
   \caption{Effective permittivity of a Ag/TiO$_2$ multilayer structure with a silver filling fraction of $f = 0.6$. The vertical
lines mark the edges of the hyperbolic bands which coincide with the epsilon-near-zero and the epsilon-near-pole points. \label{Fig:EMT}}
\end{figure}

In the following we will show that by using hyperbolic metamaterials we can use Rabi splitting for sensing of very small concentrations of atoms or molecules. To this end, we consider the Kretchmann configuration as depicted in Fig.~\ref{Fig:Kretschmann}. The prism is assumed to be made of silica with refractive index $n = 1.5$. The plasmonic film is assumed to be a hyperbolic material of thickness $d = 50\,{\rm nm}$ which is realized by a multilayer structure of alternating layers of silver and TiO$_2$ using the same material parameters as in Ref.~\cite{SABGSA2015}. The effective medium response of such a multilayer material is that of a uni-axial medium 
with permittivity
\begin{equation}
  \boldsymbol{\epsilon} = \begin{pmatrix} \epsilon_\perp & 0 & 0 \\ 0 & \epsilon_\perp & 0 \\ 0 & 0 & \epsilon_\parallel \end{pmatrix},
\end{equation}
where $\epsilon_\parallel$ is the permittivity along the optical axis which is parallel to the surface normal and $\epsilon_\perp$
is the permittivity perpendicular to the optical axis. In Fig.~\ref{Fig:EMT} the real values of both permittivities  $\epsilon_\perp$
and $\epsilon_\parallel$ are shown for a volume filling fraction of $f = 0.6$ of silver. The two hyperbolic frequency bands
of type I ($\epsilon_\parallel < 0$ and $\epsilon_\perp > 0$) and of type II ($\epsilon_\parallel > 0$ and $\epsilon_\perp < 0$) can be nicely seen. The edges of the two bands are determined by the epsilon-near-zero (ENZ) $\lambda_{\rm ENZ} = 414\,{\rm nm}$ and the epsilon-near-pole (ENP) $\lambda_{\rm ENP} = 513\,{\rm nm}$ wavelengths. It could already be shown that large 
dipole-dipole interactions are achievable close to these wavelengths~\cite{SABGSA2015,DesmukhEtAl2017}. One can expect to have a large Rabi splitting close to ENP and ENZ wavelengths as well.

To test this hypothesis we consider a thick layer of dye molecules of rhodamin 6 (R6G) in ethylene glycol solution on a hyperbolic medium as substrate as depicted in Fig.~\ref{Fig:Kretschmann}. The resonance wavelength of the R6G molecules is around $532\,{\rm nm}$ which closely matches the resonance ENP wavelength. Hence, we can expect to have a very effective coupling and Rabi splitting in this case. In order to determine the reflectivity in the Kretschmann configuration in Fig.~\ref{Fig:Kretschmann} we use for the permittivity of R6G the model~\cite{Bojarski1997} 
which is given by Eq.~(\ref{Eq:PermittivityMolecule}) with
\begin{equation}
  \omega_p^2 = C h \frac{e^2}{m \epsilon_0}
\end{equation}
%
%
%
%
with $\omega_0 = 3.5\times10^{15}\,{\rm rad/s}$, $\gamma  = 2.07\times10^{14}\,{\rm rad/s} $, $h = 0.74$. The concentration $C$ of rhodamin dye molecules is given in units of $M = {\rm mol}/{\rm l} = 1000 {\rm mol}/{\rm m}^3$. Therefore a concentration of 0.002M corresponds to 1.2 molecules in a volume of (10nm)$^3$ or a number density $n$ of $1.2\times10^{24}\,{\rm m}^{-3}$ and 0.1M corresponds to 60 molecules in a volume of (10nm)$^3$ or a number density of $6\times10^{25}\,{\rm m}^{-3}$.

 \begin{figure}[!htb]
   \includegraphics[width=0.45\textwidth]{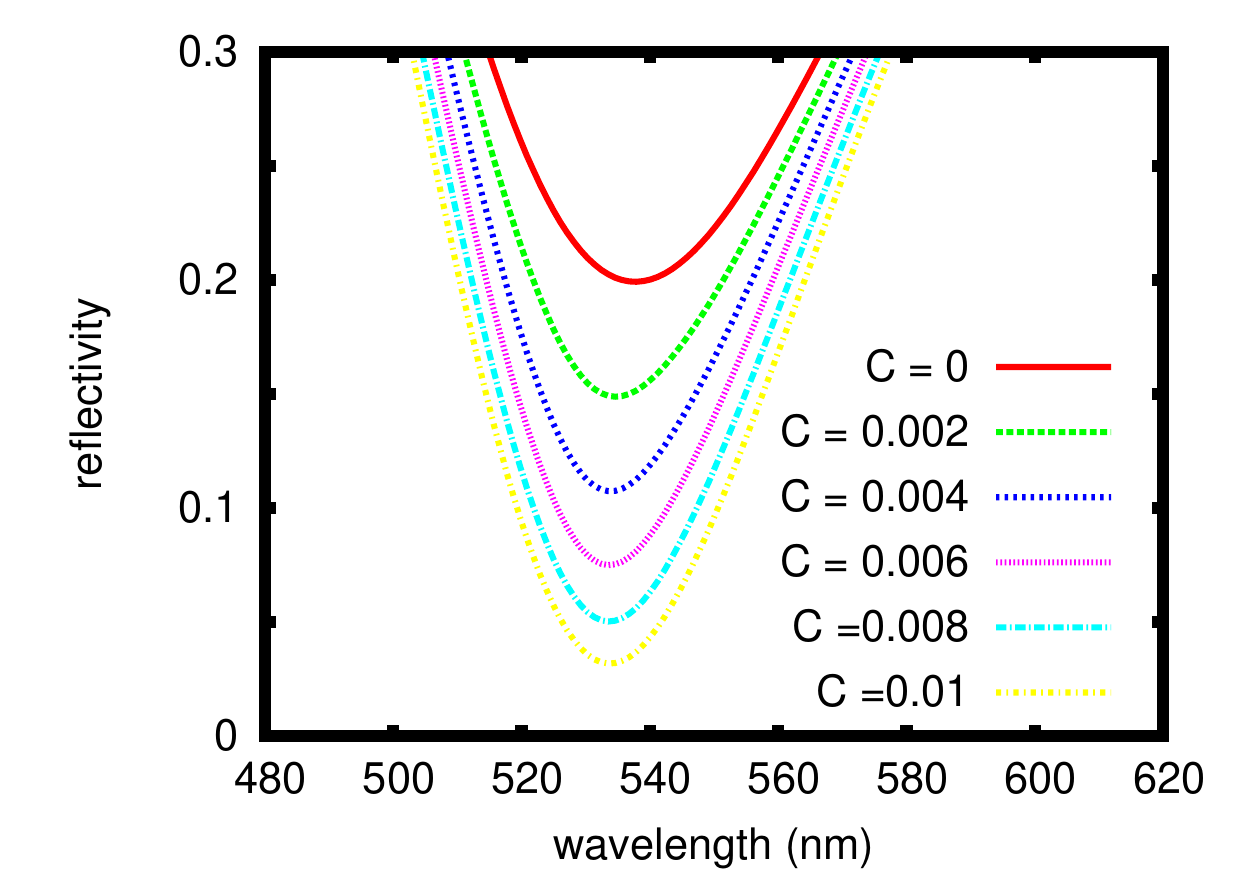}
   \includegraphics[width=0.45\textwidth]{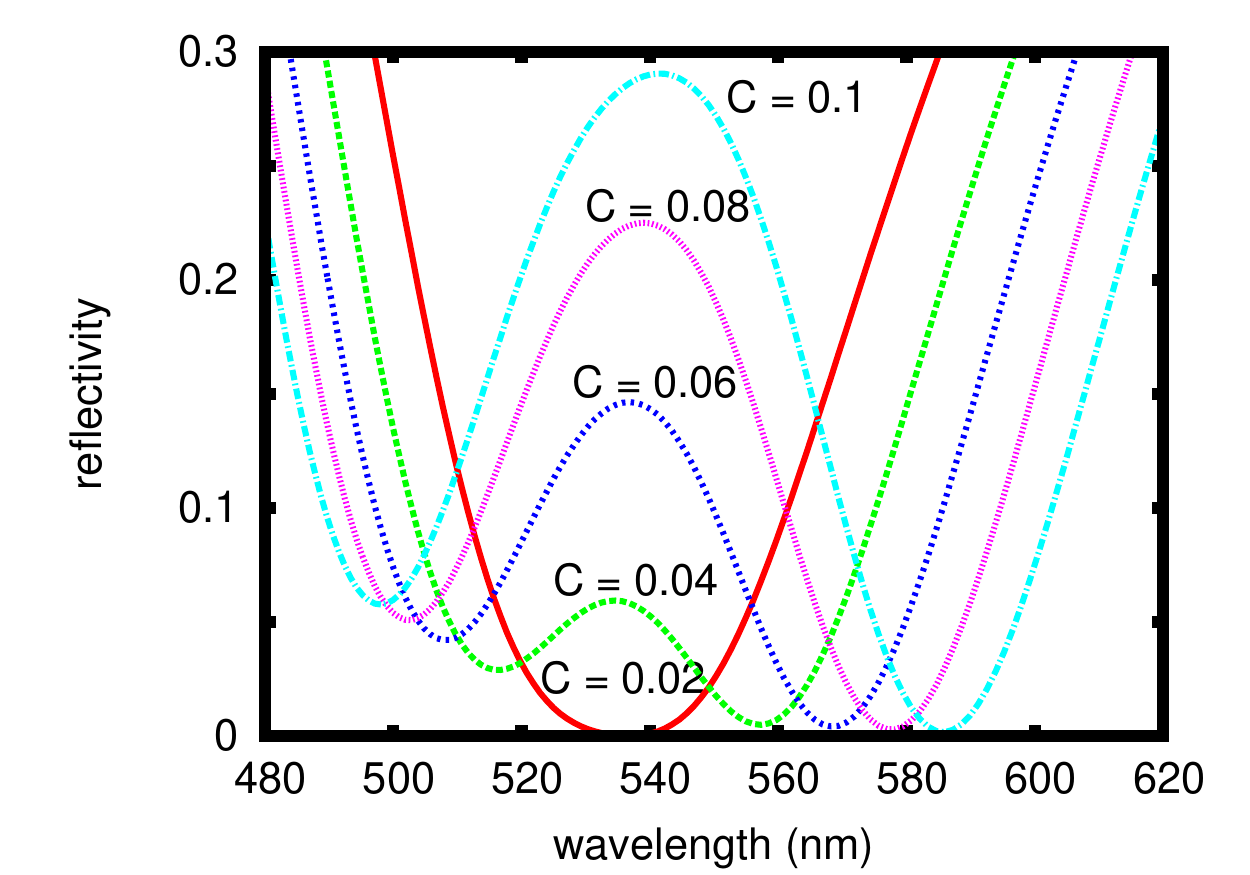}
   \caption{Reflectivity of p-polarized light in the Kretschmann configuration for different concentrations $C$ of R6G molecules. \label{Fig:Reflectance1and2}}
\end{figure}
 \begin{figure}[!htb]
   \includegraphics[width=0.45\textwidth]{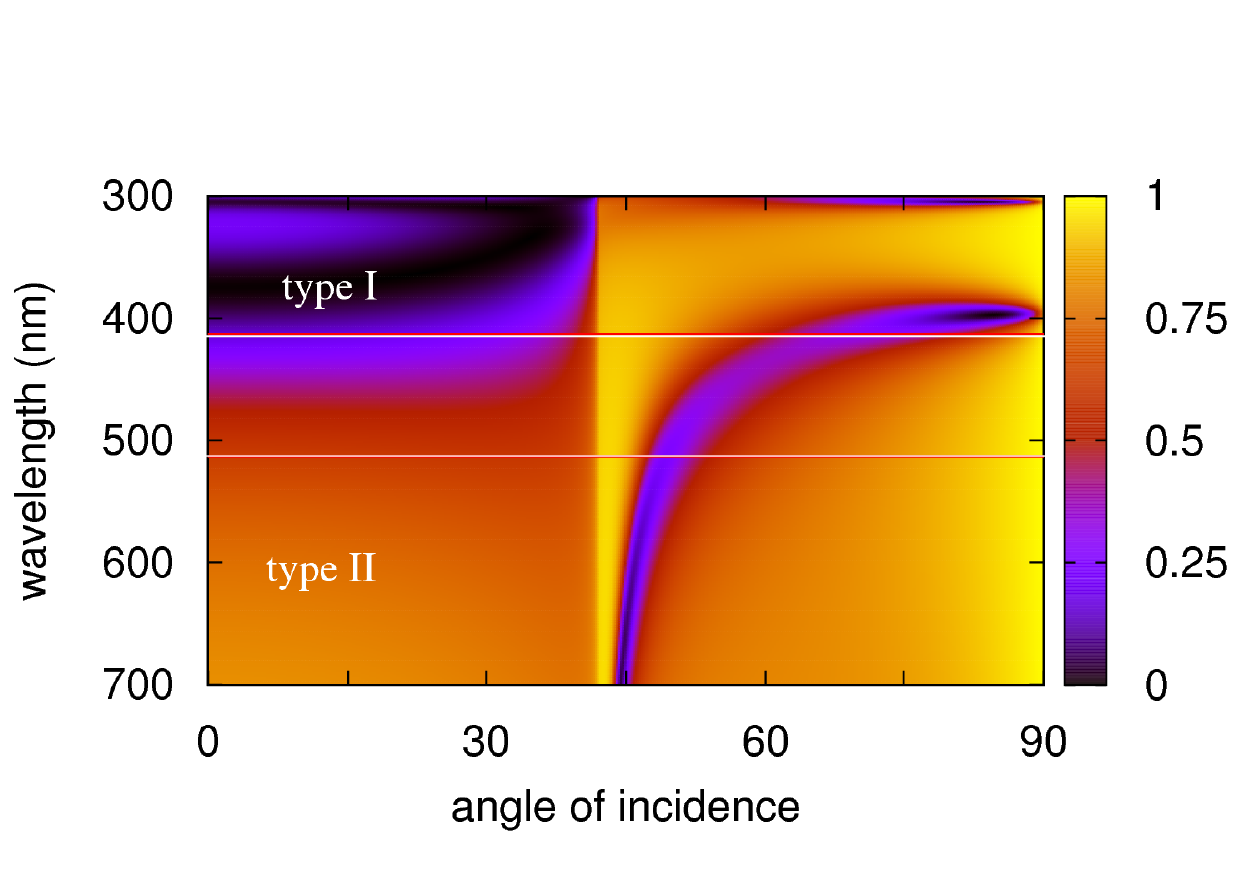}
   \includegraphics[width=0.45\textwidth]{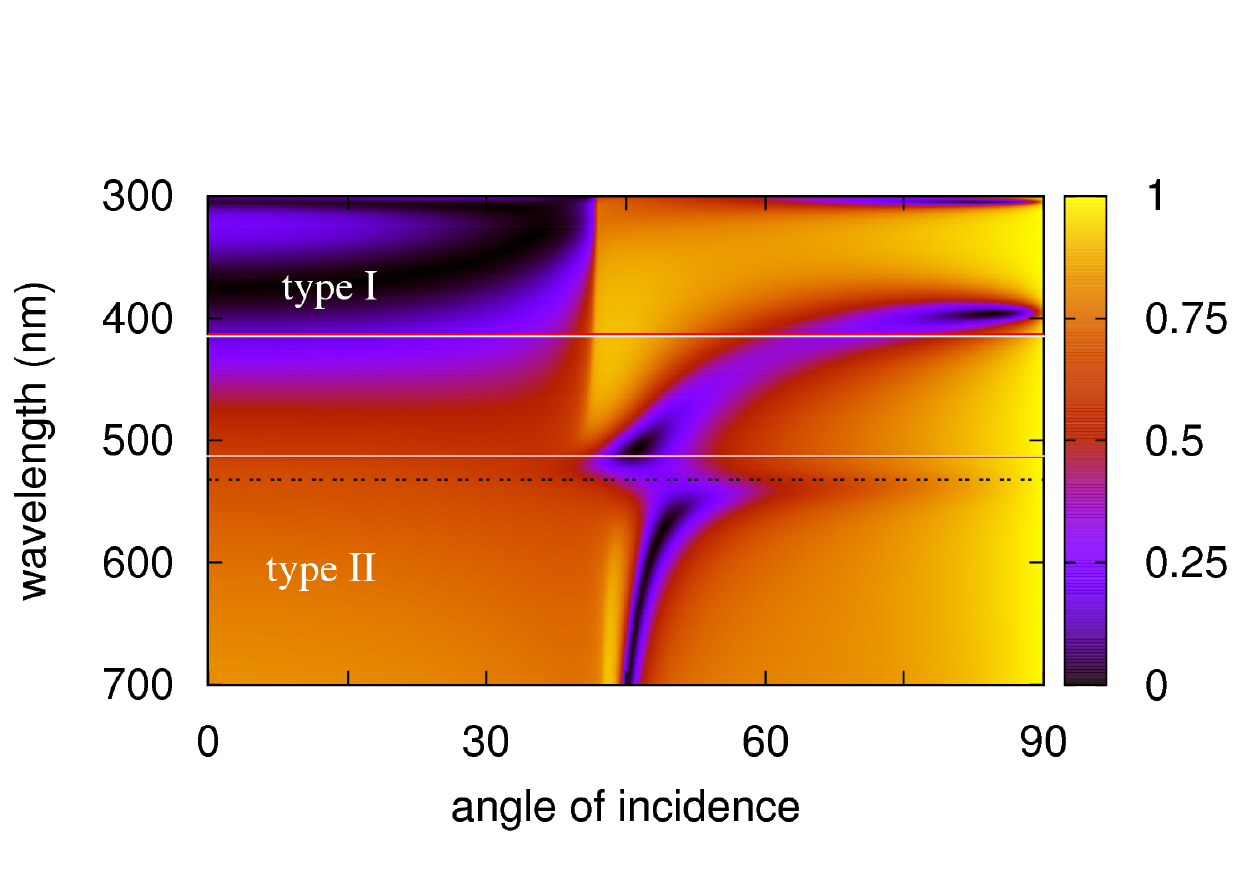}
   \caption{Reflectivity  of p-polarized light in the Kretschmann for different angles of incidence and wavelengths for a configuration of $C = 0\,{\rm M}$ or $n = 0\times10^{27}\,{\rm m}^{-3}$ (top) and  $C = 0.1{\rm M}$ or $n = 6\times10^{25}\,{\rm m}^{-3}$ (bottom) of R6G molecules. The vertical dashed black line is positioned at the resonance wavelength of R6G molecule at $532\,{\rm nm}$. The vertical solid white are positioned at the ENZ and ENP wavelength.\label{Fig:Reflectance1and2omega}}
\end{figure}

\section{Transition from weak to strong-coupling regime}

The reflectivity of p-polarized light~\cite{YeeBook} for an angle of incidence of $\theta = 48^\circ$ is shown in Fig.~\ref{Fig:Reflectance1and2} for different concentrations of R6G using effective medium theory for the hyperbolic material. It can be seen that for very low concentrations in the range of 0M-0.01M ($0{\rm m}^{-3}-6\times10^{24}\,{\rm m}^{-3}$) the minimum in the reflectivity shifts from $0.2$ to $0.04$ which is easily measurable. {Note, that there is a small blue shift. We find that for smaller angles like  $\theta = 47.5^\circ$ we have larger blue-shift, whereas for larger angles like $\theta = 49^\circ$ we find a red-shift. Therefore the reflection dip has the tendency to shift towards $\omega_0$.}

For large concentrations of 0.02M-0.1M ($1.2\times10^{25}\,{\rm m}^{-3}-6\times10^{25}\,{\rm m}^{-3}$) the Rabi splitting sets in. Again by measuring only the shift of the small wavelength minimum one can easily measure the concentration of the dye molecules. Note that the separation between the two dips at $498\,{\rm nm}$ ($2.49\,{\rm eV}$) and $586\,{\rm nm}$ ($2.12\,{\rm eV}$) for a high concentration of 0.1M ($6\times10^{25}\,{\rm m}^{-3}$) is $\Delta \lambda = 88\,{\rm nm}$ which corresponds to 370meV. Let us compare this value with the estimate given in Eq.~(\ref{Eq:Estimate}). For the R6G molecule at the highest concentration we have $\omega_p = 3.76\times10^{14}\,{\rm rad/s} \ll \omega_0 = 3.5\times10^{15}\,{\rm rad/s}$ and $\omega_p > \gamma = 2.1\times10^{13}\,{\rm rad/s}$. Therefore, from the relation (\ref{Eq:Estimate}) a Rabi splitting of about 247meV can be expected. This is in agreement with the experimentally found values of $100-200\,{\rm meV}$ observed in \cite{Hakala2009}. The here observed value of 370meV with the hyperbolic film is larger than the estimate for an ideal perfect cavity which is due to the detuning between the ENZ frequency which corresponds to $\omega_c$ and the resonance frequency $\omega_0$ of the dye molecule {and the anisotropic character of the hyperbolic medium}.

An angle dependent plot of the reflectivity in p-polarization for this the given Kretschmann-Raether configuration is shown in Fig.~\ref{Fig:Reflectance1and2omega} in presence of 0.1M  ($6\times10^{25}\,{\rm m}^{-3}$) of R6G in comparison to the case where no dye molecules are present in Fig.~\ref{Fig:Reflectance1and2omega}. The Rabi strong coupling between the dye molecules and the resonance of the hyperbolic metamaterial can be nicely seen. Obviously the Rabi splitting can be found in a range of angles of incidence of about 45$^\circ$ to 55$^\circ$, so that the effect is relatively robust with respect to variations in the angle of incidence.

\section{Conclusion}

In summary, we have shown that by using hyperbolic metamaterials in a Kretschmann-Raether configuration one can measure even small concentrations of R6G molecules of only a few mM. For large concentrations one can easily observe the transition from the weak coupling to the strong coupling regime. The Rabi splitting is directly observable in the reflectivity signal. Therefore hyperbolic materials are also intersting platforms for Rabi splitting measurements as well as sensing applications. Finally, hyperbolic materials can also serve as a platform for studying strong coupling phenomena in J-aggregates and two-dimensional transition metal dichalcogenides like MoS$_2$, which are promissing candidates for room temperature exitonic devices in opto-electronics, photovoltaics and ultra-fast photodetection~\cite{LundtEtAl2017}.  

%
%
%

\section*{acknowledgments}

The authors thank the Bio Photonics initiative of the Texas A \& M university for supporting this work. S.A.B.~thanks the hospitality of the Texas A \& M university.

\end{document}